\documentclass[preprint, 10pt, 5p]{elsarticle}
\usepackage[utf8]{inputenc}
\usepackage{geometry} 
\usepackage{graphicx} 
\usepackage{amsmath, amsbsy}
\usepackage{amssymb}
\usepackage{float} 
\usepackage{wrapfig} 
\usepackage[hidelinks]{hyperref}
\usepackage{tikz}

\biboptions{numbers, sort&compress}

\begin{document}
\title{Trace Impurity Transport in Multi-Species Plasmas with Large Particle Fluxes}
\date{\today}
\author[hunteraddress]{E. Litvinova Mitra}
\ead{elena.mitra49@myhunter.cuny.edu}
\author[princetonaddress]{E. J. Kolmes}
\author[princetonaddress]{I. E. Ochs}
\author[princetonaddress]{M. E. Mlodik}
\author[princetonaddress]{T. Rubin}
\author[princetonaddress]{N. J. Fisch}
\address[hunteraddress]{Department of Physics and Astronomy, Hunter College, New York, NY 10065, USA}
\address[princetonaddress]{Department of Astrophysical Sciences, Princeton University, Princeton, NJ 08544, USA}

\begin{abstract}
A burning fusion plasma has a large inwards flux of fuel and outwards flux of ash. Existing impurity transport theories do not account for a steady-state environment of such large fluxes. In this paper, we extend classical transport theory to account for a background of such fluxes. In a mainly two-ion species magnetized plasma, with oppositely directed density gradients maintained by oppositely directed sources of the two species, trace impurity ions are shown to align with the larger mass ions. This unexpected behavior is derived analytically and simulated numerically using the MITNS (Multiple-Ion Transport Numerical Solver) code. The result suggests a potential advantage of edge-fueled p-B11 fusion in extracting high-Z impurities, and possible applications in plasma-based separation methods. 
\end{abstract}

\begin{keyword}
Trace impurity \sep impurity transport \sep magnetized plasma \sep particle flux \sep multiple species plasma
\end{keyword}

\maketitle

\section{Introduction\label{sec:Intro}}
Cross-field transport of multiple ion species is a complex and important phenomenon in plasma science \cite{Spitzer1952,Taylor1961ii,Braginskii1965,Connor1973,Rutherford1974,Hinton1974}. Its applications range widely and include controlled fusion devices, such as tokamaks \cite{Hirshman1981,Redi1991,Fisch1992,Wade2000,Dux2004,Hole2014, Zonca2014}, stellarators \cite{McCormick2002,Braun2010,Helander2017,Newton2017}, and magnetized inertial fusion devices, such as MagLIF \cite{Slutz2010,Slutz2012,Ochs2018PRL}. Furthermore, non-fusion devices such as plasma mass filters \cite{Bonnevier1966,Lehnert1971,Hellsten1977,Krishnan1983,Geva1984,Prasad1987,Bittencourt1987,Grossman1991,Fetterman2011,Ochs2017iii,Dolgolenko2017,Zweben2018,Gueroult2018} and non-neutral particle traps \cite{Davidson1970,ONeil1979,Prasad1979,ONeil1981,Imajo1997,Dubin1999} also utilize cross-field transport predictions in separating plasma components.

In classical magnetized plasma transport theory \cite{Braginskii1965, Taylor1961ii, Spitzer1952}, on the fast ion-ion transport time-scale, and, when all gradients are in the $\hat{x}$ direction, perpendicular to the magnetic field, $\mathbf{B}$, in the $\hat{z}$ direction, with $\hat{y}$ an ignorable direction, as in the schematic in Fig.~\ref{fig:FIG1_geometry}, the ion species in the plasma will arrange themselves according to:  
\begin{gather}
n_a^{1/Z_a} \propto n_b^{1/Z_b}, \label{eqn:impurityPinch}
\end{gather}
where $n_i$ and $Z_i$ are the density and charge state of ion species $i$ in $(a,b)$, and where we do not consider the effects of electron transport, which is on relatively longer time-scales than ion-ion transport time-scale. This relation can be generalized in the presence of a species-dependent external potential $\Phi_i$ \cite{Kolmes2018,Kolmes2020MaxEntropy}:
\begin{gather}
\big( n_a e^{\Phi_a/T} \, \big)^{1/Z_a} \propto \big( n_b e^{\Phi_b/T} \, \big)^{1/Z_b} . \label{eqn:generalizedPinch}
\end{gather}
Incidentally, this generalization also shows how the centrifugal forces in rotating plasma can be used to  exclude impurities, including ash in the case of nuclear fusion from the plasma core, even as essential ions are retained \cite{Kolmes2018,Kolmes2020MaxEntropy}.

Eq.~(\ref{eqn:generalizedPinch}) assumes that the ions relax to a flux-free steady state. However, in plasma experiments and applications, we are often examining a driven steady state. For example, in a fusion reactor, fuel is fed into the system at the edges, and there is an ash flux from the core. It is important to understand whether these fluxes affect the impurity accumulation \cite{Ochs2018_anisotropy,Knapp2019,Gomez2019,Schmit2014}. 

In this paper, we derive the density profiles in the presence of a steady-state particle flux. We confirm our analytic results numerically using the MITNS code \cite{Kolmes2020MITNS}, in three scenarios: in the presence of two ion sources and sinks and no external potential, Sec.~\ref{sec:Experiment1_no_potential}; in the presence of a non-zero external potential, Sec.~\ref{sec:Experiment2_periodic_potential}; as well as in the presence of a trace impurity but without an external potential, Sec.~\ref{sec:Experiment3}. In doing so, we find that the presence of steady-state particle flows can substantially alter the behavior of Eqs.~(\ref{eqn:impurityPinch}) and (\ref{eqn:generalizedPinch}). For example, in the case of two counter-flowing bulk ion species, we show that a trace third species' density gradient aligns with that of the larger mass background ion species.

This key result reveals an advantage of edge-fueled pB-11 fusion over DT fusion since the impurities will be extracted as they try to align with the high-mass Boron fuel that flows in from the edge. Our results also inform on the transport dynamics in plasma-based mass filtering methods. These applications are discussed in Sec.~\ref{sec:Discussion_and_Conclusion}.

\section{\label{sec:Method}Method}

\begin{figure}
\includegraphics[width=\linewidth]{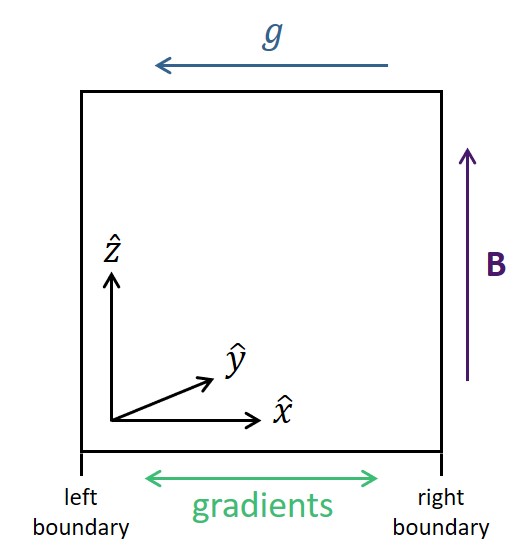}\caption{\label{fig:FIG1_geometry}This schematic shows the basic geometry used in this paper.}
\end{figure}

To examine the driven steady state, we use models of classical transport throughout this paper. These models are valid only when the plasma is sufficiently quiescent. Turbulent transport may dominate in regimes where significant fluctuations are excited, and factors such as density and temperature gradients can become important \cite{Horton1999}. Thus, the governing classical transport equations for continuity and momentum \cite{Zhdanov,Kolmes2020MITNS} are modified by including a particle source term $s_i$ for species $i$ as follows:
\begin{equation}
\frac{\partial n_{i}}{\partial t}+\nabla\cdot(n_{i}\mathbf{v}_{i})=s_{i}\label{eq:Continuity}
\end{equation}
\begin{multline}
\frac{\partial}{\partial t}(m_{i}n_{i}\mathbf{v}_{i})+\nabla p_{i}+\nabla\cdot(\pi_{i}+m_{i}n_{i}\mathbf{v}_{i}\mathbf{v}_{i})\\
=Z_{i}en_{i}(\mathbf{E+}\mathbf{v}_{i}\times\mathbf{B\mathrm{)}}+m_{i}n_{i}\sum_{s}\nu_{is}(\mathbf{v}_{s}-\mathbf{v}_{i})\\
+\mathbf{f}_{th,i}+m_{i}n_{i}\mathbf{g}+m_{i}s_{i}\mathbf{v}_{i}^{src}\label{eq:Momentum},
\end{multline}
where $\mathbf{v}_{}$ is the ion fluid velocity; $p_i$ is the scalar pressure; $\pi_i$ is the viscosity tensor; $e$ is the elementary charge; $Z_i$ is the ion charge state; $\mathbf{E}$ is the electric field; $\mathbf{B}$ is the magnetic field; $m_i$ is the ion mass; $\nu_{is}$ is the collision frequency of species $i$ with species $s$; $\mathbf{g}(t,x)$ is the gravitational acceleration; $\mathbf{f}_{th,i}$ is the cross-field thermal force density on species $i$ \cite{Kolmes2020MITNS} defined as: 
\begin{multline}
\mathbf{f}_{th,a}=\sum_{b}\mathop\frac{3}{2}\frac{n_{a}\nu_{ab}}{\Omega_{a}}\frac{1}{1+(m_{a}T_{a}/m_{b}T_{b})}\\
\times\left(\hat{b}\times\nabla T_{a}-\frac{Z_{a}}{Z_{b}}\frac{m_{a}}{m_{b}}\frac{T_{b}}{T_{a}}\hat{b}\times\nabla T_{b}\right),
\end{multline}
where the characteristic proton gyrofrequency is $\Omega_{a}\doteq{eB_{0}/m_{p}}$ for proton mass $m_p$. The system constitutes a slab of magnetized plasma with boundaries on the $x$-axis, $\mathbf{B}$ in the $\hat{z}$ direction, and all gradients along $\hat{x}$ as shown in Fig.~\ref{fig:FIG1_geometry}.
In the classical transport model, we do not consider electron transport since electron flux is much slower than ion flux, and we assume that, in a steady state, we can control the overall density of the electrons. However, in non-classical regimes where turbulence dominates, the electron flux could be significant.

To derive the flux for any two species $a$ and $b$ with constant sources and sinks, define the flux of species $a$ as $\Gamma_{a}=n_a\mathbf{v}_{a\hat{x}}$. Define $\Gamma_{ab}$ to be the contribution to the flux due to collisions with species $b$. The source term and the flux are related by the continuity equation, Eq.~(\ref{eq:Continuity}), as: $\partial{\Gamma_{ab}}/\partial{x}=s_i$.
When the time derivative vanishes; from the assumption of classical transport the momentum advection and viscosity are neglected; and the frictional force is only in the direction of the gradients, it follows directly from Eq.~(\ref{eq:Momentum}) that \cite{Kolmes2018}:
\begin{multline}
\Gamma_{ab}=-\frac{\tilde{\nu}_{ab}}{\Omega_{a}}\frac{T}{B}n_{a}n_{b}\\
\times(\frac{1}{Z_{a}}\frac{n_{a}^{'}}{n_{a}}+\frac{1}{Z_{a}}\frac{\Phi_{a}^{'}}{T}-\frac{1}{Z_{b}}\frac{n_{b}^{'}}{n_{b}}-\frac{1}{Z_{b}}\frac{\Phi_{b}^{'}}{T}),\label{eq:Flux}
\end{multline}
where the primes denote a partial derivative with respect to $x$; $\Phi_i$ is the external potential for ion species $i$; and ${\nu}_{ab}\doteq{\tilde{\nu}_{ab}}{n_b}$. Note, when $\Gamma_{ab}$ vanishes, the density relation is the known result in Eq.~(\ref{eqn:impurityPinch}). We, however, maintain a finite flux term. Let $\zeta=Z_{a}/Z_{b}$ and rearrange terms to get:
\begin{multline}
[n_{a}n_{b}^{-\zeta}e^{[\Phi_{a}-\zeta\Phi_{b}]/T}]^{'}\\
=\frac{-Z_{a}e\Omega_{a}B\Gamma_{ab}}{T\tilde{\nu}_{ab}}n_{b}^{-\zeta-1}e^{[\Phi_{a}-\zeta\Phi_{b}]/T}\label{eq:nanb}.
\end{multline}
The above expression is a differential equation describing the generalization of the pinch to finite flux. In subsequent sections, we examine several specific cases to understand its implications.

\subsection{\label{sec:Experiment1_no_potential}Case 1: Two sources and no potential}
In our first case, two source are added at either boundary (see Fig.~\ref{fig:FIG1_geometry}) with no potential acting on the system, so we set $\Phi=0$, and integrate Eq.~(\ref{eq:nanb}) to get the following:
\begin{equation}
n_{a}n_{b}^{-\zeta}-n_{a_0}n_{b_0}^{-\zeta}=-\frac{Z^{2}_a e^{2} {B}^{2}}{{m}_{a}{T}\tilde{\nu}_{ab}}\intop_{x_0}^{x}\Gamma_{ab}n_{b}^{-\zeta-1}dx,
\end{equation}
were $x_0$ is a reference point, and $n_{a_0}$, $n_{b_0}$ are ion densities at $x_0$. Then solving for the density of species $a$, we have:
\begin{equation}\label{eq:nopotential}
{n}_{a}=n_{b}^{\zeta}\Bigg({n}_{a_0}{n}_{b_0}^{-\zeta}-\frac{Z^{2}_a e^{2}{B}^{2}}{{m}_{a}{T}{\tilde{\nu}_{ab}}}\intop_{x_0}^{{x}}{\Gamma}_{ab}{n}_{b}^{-\zeta-1}d{x}\Bigg).
\end{equation}
This key result is a generalization of Eq.~(\ref{eqn:impurityPinch}) for two species in the presence of a finite steady-state flux. We seek to both verify it and to explore its implications.

\begin{figure}
\includegraphics[width=\linewidth]{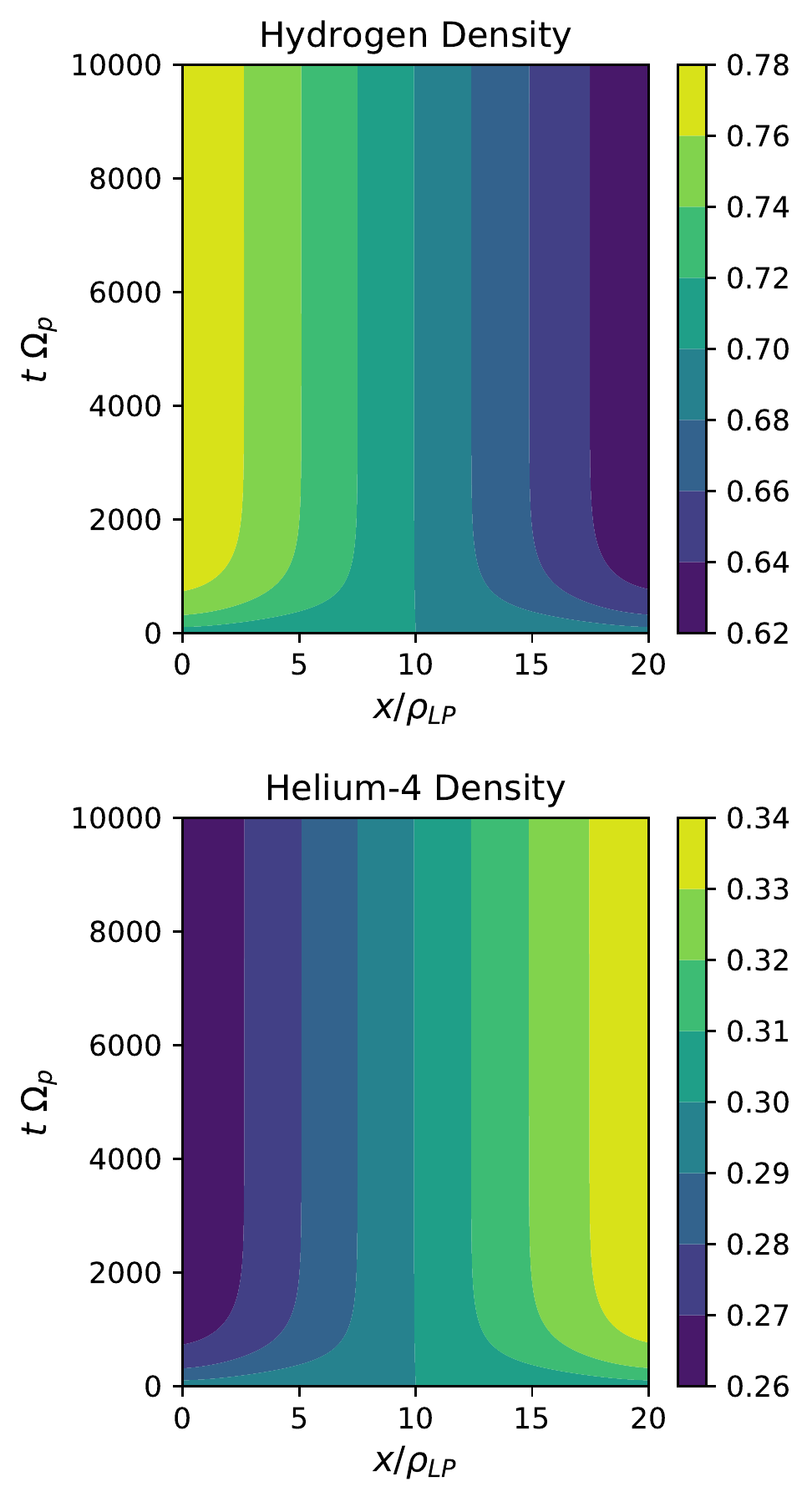}
\caption{\label{fig:FIG2_countour_gMax0} Contour density plot of hydrogen and helium-4 over $10,000$ gyroperiods. The hydrogen source is on the left (with relative density $n_0=0.7$), corresponding to the highest density, and the sink is on the right. The helium-4 source is on the right ($n_1=0.3$) with the sink on the left. Without a potential acting on the system, the density distribution remains aligned with the respective sources and sinks.}
\end{figure}
\begin{figure}
\includegraphics[width=\linewidth]{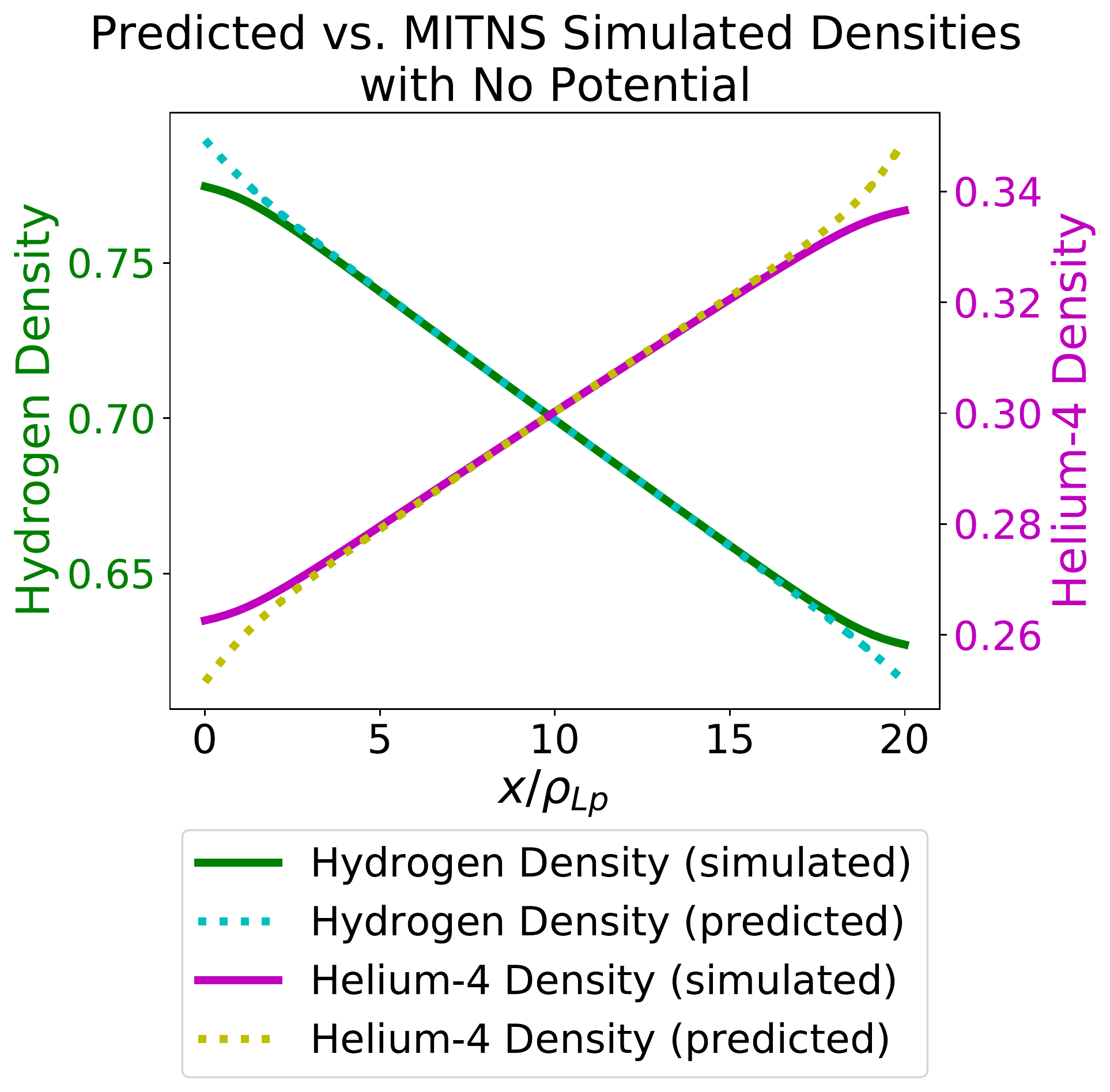}
\caption{\label{fig:FIG3_DensityComp_gMax0} A comparison between MITNS-simulated and analytically predicted densities for source ion species hydrogen and helium-4 at equilibrium with source factor $0.001$, and arbitrary units. Since our sources are volumetric, we do not expect perfect agreement at the boundaries.}
\end{figure}
We verify this result numerically in MITNS by including source and sink terms in the code. Our setup mirrors that of the original MITNS code publication \cite{Kolmes2020MITNS}: the plasma slab geometry is as in Fig.~\ref{fig:FIG1_geometry} with boundaries on the $x$-axis. The sources are added with a specified source rate at either boundary of the slab and a sink at the opposite boundary. We define the source function as: 
\begin{equation}
{s_i}=\frac{S_i}{L_i}\Big(e^{-\frac{x}{L_i}}-{e}^{-\frac{L-x}{L_i}}\Big),
\end{equation}
where $S_i=-(Z_i/Z_s)S_s$, $L_i$ is the source scale length, $L$ is the total length of the system, and $x$ is an arbitrary position along the $x$-axis. The shape of the source function is exponential, which allows sources to be treated volumetrically, and will mimic boundary sources' behavior so long as $L_i\ll L$. In other words, when $L_i/L$ is small, the analytic and simulated solutions are equivalent in the interior region of the plasma slab, since the localized volumetric sources fix the total particle fluxes in steady state. The codes' initial conditions are such that the sources are added at zero velocity, and temperature evolution is turned off so that $T$ is treated as a fixed quantity for all species.

As an example, Fig.~\ref{fig:FIG2_countour_gMax0} shows a MITNS simulation with no potential, $70\%$ hydrogen and $30\%$ helium-4, 128-cell grid and a system size of $20$ proton gyroradii. The error tolerance is set to $10^{-14}$, with initial conditions such that the plasma $\beta=10^{-3}$, and the ratio of the proton-proton collision frequency to the proton gyrofrequency is set to $0.05$. The total running time for the system is set to $10,000$ proton gyroperiods. Hydrogen ions are added at the left
boundary ($x=0$), and helium ions are added at the right
boundary ($x=20$). These parameters are used for all simulations henceforth unless stated otherwise. 

As expected, the two ion species densities show a gradient consistent with the side at which they are
added; thus, hydrogen density is highest near the left boundary,
and helium-4 density is highest at the right boundary.

Fig.~\ref{fig:FIG3_DensityComp_gMax0} shows that our simulations are in good agreement with the analytically predicted densities in Eq.~(\ref{eq:nopotential}) of species $a$ and $b$ at equilibrium and away from the boundaries.
\subsection{\label{sec:Experiment2_periodic_potential}Case 2: Non-zero potential}
\begin{figure}
\includegraphics[width=\linewidth]{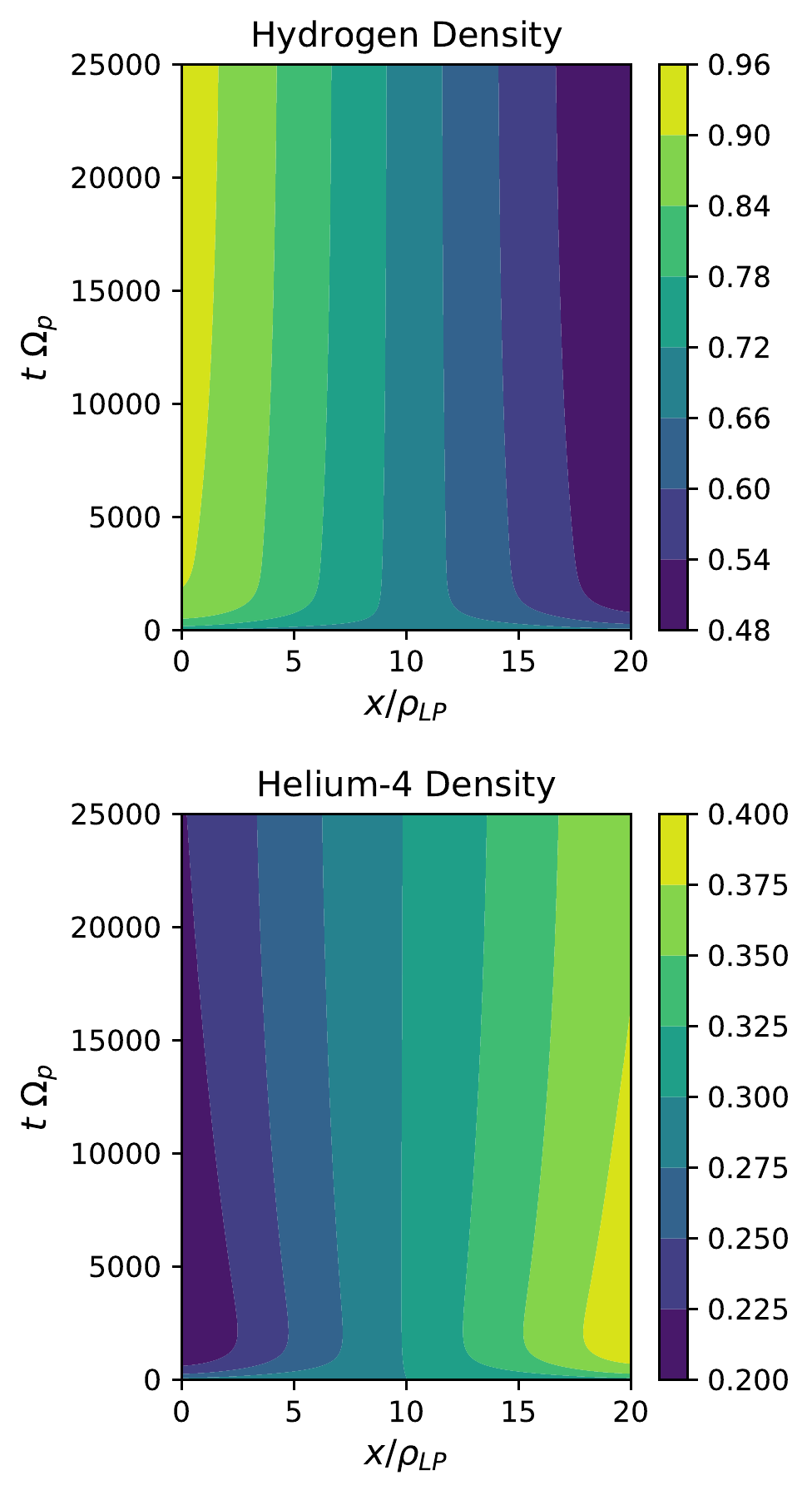}
\caption{\label{fig:FIG4_Contour_gMax01_sf003_long} Contour density plot of hydrogen (relative density $0.7$) and helium-4 ($0.3$) as in the no potential case. The $t$ is set to $25,000$ gyroperiods, and a potential with a maximum of $10\%$ of the product of the gyrofrequency and the proton thermal velocity ramp up time is 100 gyroperiods.}
\end{figure}
\begin{figure}
\includegraphics[width=\linewidth]{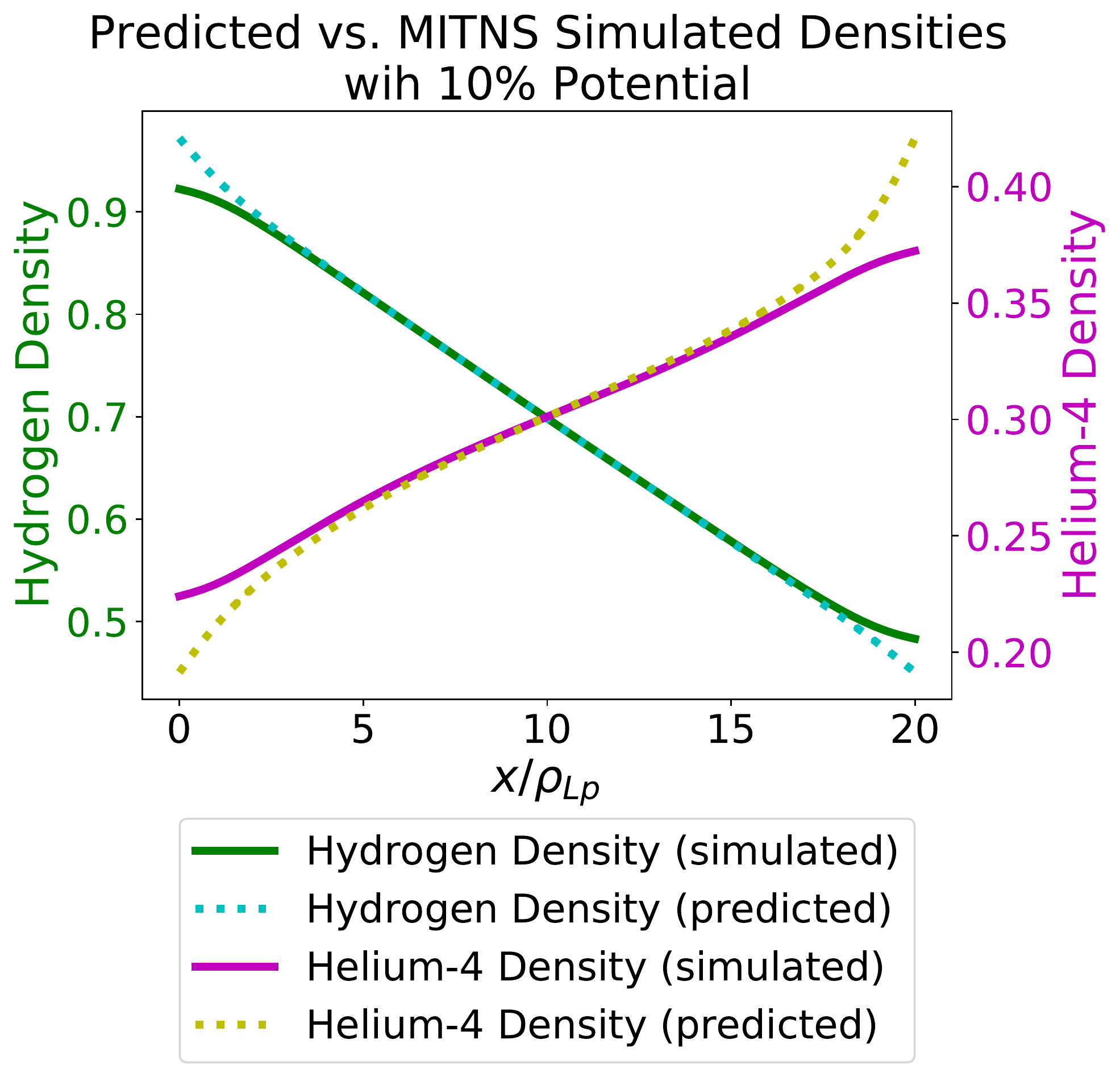}
\caption{\label{fig:FIG5_DensityComp_gMax01_sf003_long} The density of the hydrogen and helium-4 species, with arbitrary units, computed from the predicted result of Eq.~(\ref{eq:potential_n_a}) and plotted against the simulated result, with source factor $0.003$, and $10\%$ maximum potential. We expect the results to agree away from the boundaries.}
\end{figure}
In real world applications and experiments, an external potential in the form of a centrifugal or a gravitational force may be acting on the system. We derive and simulate this scenario here. From Eq.~(\ref{eq:nanb}), the density of species $a$ is:
\begin{multline}
n_a=n_b^{\zeta}e^{-[\Phi_{a}-\zeta\Phi_{b}]/T}\Biggl({n}_{a_0}{n}_{b_0}^{-\zeta}e^{[\Phi_{a_0}-\zeta\Phi_{b_0}]/T}\\
-\frac{Z^{2}e^{2}{B}^{2}}{{m}_{a}{T}{\tilde{\nu}_{ab}}}\intop_{x_0}^{{x}}{\Gamma}_{ab}{n}_{b}^{-\zeta-1}e^{[\Phi_{a}-\zeta\Phi_{b}]/T}dx\Biggr)\label{eq:potential_n_a},
\end{multline}
where $\Phi_{a_0}$ and $\Phi_{b_0}$ are the potentials for species $a$ and $b$ at $x_0$. This is a finite-flux generalization of Eq.~(\ref{eqn:generalizedPinch}). We then choose a potential ${\Phi(t,x)}$ defined as:
\begin{equation}
\frac{\Phi(t,x)}{T}=-\frac{m_{i}g_{o}L}{\pi T}\tanh^4(\frac{t}{t_{ramp}}) \cos(\frac{\pi x}{L}).
\end{equation}
Note that, at $t\gg t_{ramp}$, $\tanh^4(\frac{t}{t_{ramp}})\rightarrow1$. This analytic result predicts a similar density profile as in the no potential case: the bulk species gradient is in the direction of the particle source despite the non-zero potential. We simulate the resulting densities in Fig.~\ref{fig:FIG4_Contour_gMax01_sf003_long}, which agree with the predicted densities in Eq.~(\ref{eq:potential_n_a}) as verified in Fig.~\ref{fig:FIG5_DensityComp_gMax01_sf003_long}.

\subsection{\label{sec:Experiment3}Case 3: Two sources with a trace impurity}
\begin{figure}
\includegraphics[width=\columnwidth]{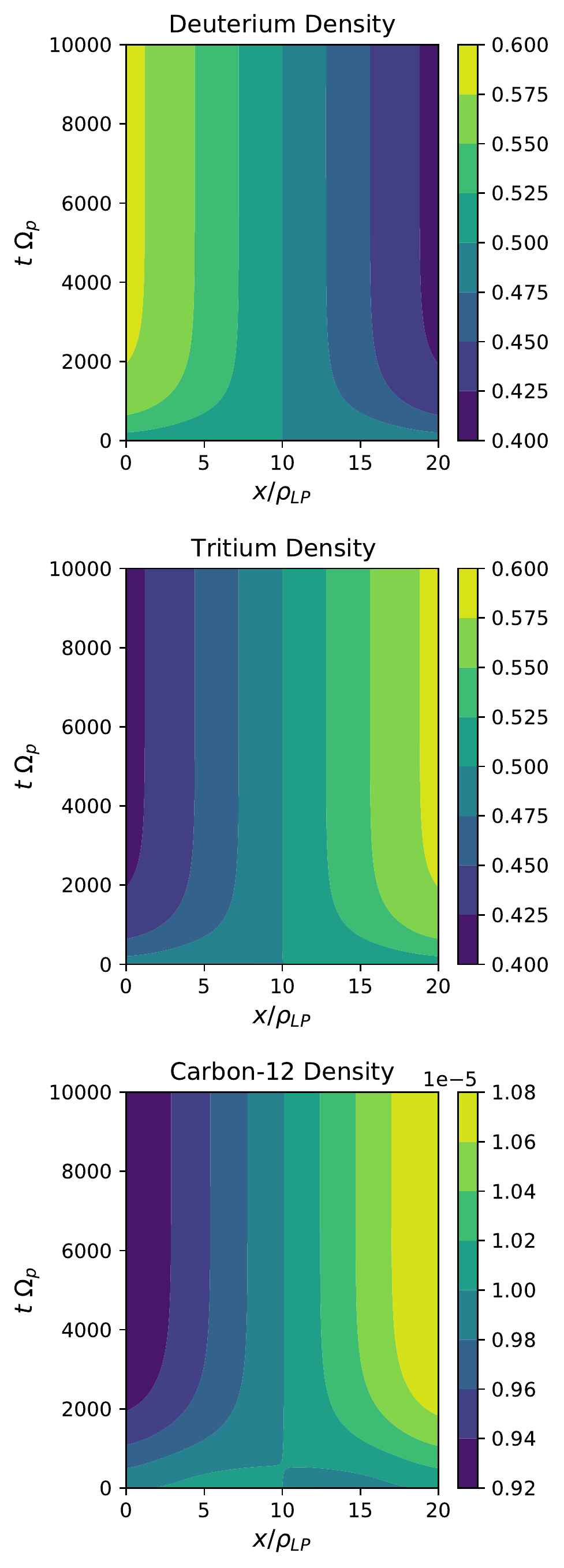}
\caption{\label{fig:FIG6_Contour_Carbon12} Contour density plot of deuterium (relative density $0.5$), tritium ($0.49999$), and carbon-12 ($10^{-5}$) over $10,000$ gyroperiods, and no potential. The sources are on either side of the slab, with deuterium on the left and tritium on the right. The sinks are on the opposite sides, respectively. Carbon-12 is a trace impurity distributed uniformly throughout the system at $t=0$. As the system approaches equilibrium, the trace shifts toward the higher mass bulk ion species, tritium, on the right.}
\end{figure}
\begin{figure}
\includegraphics[width=\linewidth]{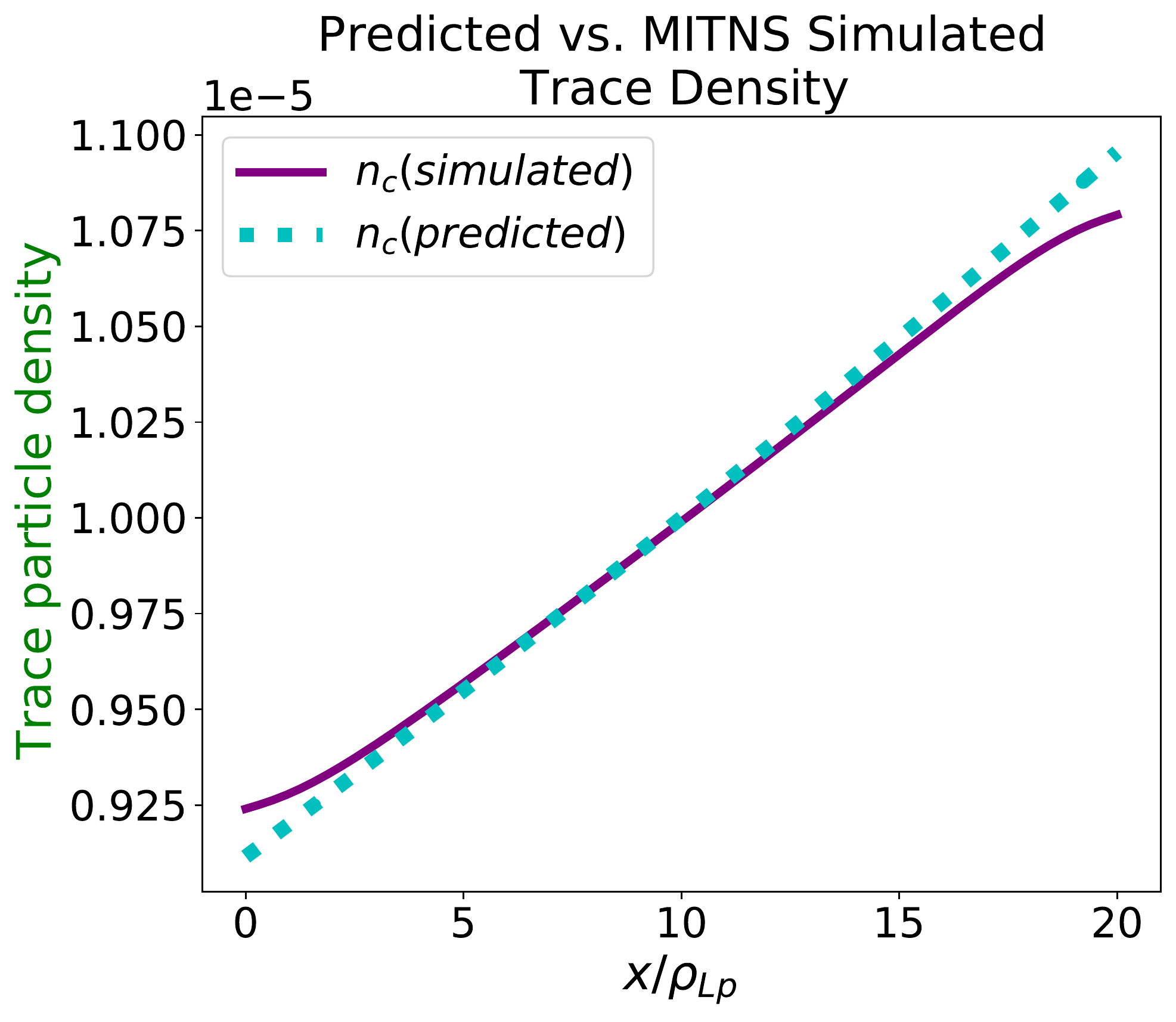}
\caption{\label{fig:FIG7_Density_Carbon12} The density of trace carbon-12 ($n_c$), in arbitrary units, with background ion sources, deuterium and tritium, computed from the predicted result using Eq.~(\ref{eq:n_c}) against the simulated result.}
\end{figure}
A fascinating case for controlled and magneto-inertial fusion is a plasma that contains a trace impurity. Both DT and pB-11 fusion produce energy by emitting charged particles, and in both cases, the fusion efficiency is sensitive to impurities coming from the reactor's wall material. Studying the impurity dynamics can help expel such impurities, thus increasing the fusion rate. In this section, we consider this scenario by examining a trace species behavior in a background cross-field flow of two ion species.

Consider a system containing two ion species $a$ and $b$, a third trace species $c$ with no potential, and a constant temperature. For a slab geometry, all gradients are in the $\hat{x}$ direction with $\mathbf{B}$ in the $\hat{z}$ direction. To determine the density profile of the trace species, we need an equilibrium relation. At equilibrium, there is no flux of trace species through the system. Therefore, there should be no net force acting on the trace species in the drift direction. The only force acting in that direction is the friction force, where the friction force density is $R_{is}$ for an arbitrary ion species $i$ and $s$. We refer to the bulk species as species $a$ and $b$, and a third species as $c$. Hence, the equilibrium condition from energy conservation \cite{Braginskii1965} is: 
\begin{equation}\label{eq:energy_conservation}
\mathbf{R}_{ca}+\mathbf{R}_{cb}=0,  
\end{equation}
where: 
\begin{align}
\begin{split}
\mathbf{R}_{ca}=m_{c}n_{c}\nu_{ca}(\mathbf{v}_{a}-\mathbf{v}_{c})\\
\mathbf{R}_{cb}=m_{c}n_{c}\nu_{cb}(\mathbf{v}_{b}-\mathbf{v}_{c}).
\end{split}
\end{align}
We can compute $\mathbf{v}$, the drift velocity in the $\hat{y}$ direction, as:
\begin{align} \label{eq:driftvel}
\begin{split}
\mathbf{v}_{i} &= \frac{\mathbf{E}\times\hat{b}}{B} -\frac{\nabla p_{i}\times\hat{b}}{m_{i}n_{i}\Omega_{i}}\\
     &=\frac{\mathbf{E}\times\hat{b}}{B}-\frac{\nabla p_{i}\times\hat{b}}{m_{i}n_{i}\left(\frac{Z_{i}eB}{m_{i}}\right)}.
\end{split}
\end{align}
Combining Eqs.~(\ref{eq:energy_conservation}) and (\ref{eq:driftvel}) we have:
\begin{multline}
m_{c}n_{c}\nu_{ca}\left(\frac{\nabla p_{c}}{n_{c}Z_{c}B}-\frac{\nabla p_{a}}{n_{a}Z_{a}B}\right)\\
+m_{c}n_{c}\nu_{cb}\left(\frac{\nabla p_{c}}{n_{c}Z_{c}B}-\frac{\nabla p_{b}}{n_{b}Z_{b}B}\right)=0.
\end{multline}
Then cancelling $B$ and $m_{c}n_{c}$ terms and substituting $\nu_{ca}\propto C_{ca}n_{a}$ and $\nu_{cb}\propto C_{cb}n_{b}$:
\begin{multline}
C_{ca}n_{a}\left(\frac{n^\prime_{c}}{n_{c}Z_{c}}-\frac{n^\prime_{a}}{n_{a}Z_{a}}\right)\\
+C_{cb}n_{b}\left(\frac{n^\prime_{c}}{n_{c}Z_{c}}-\frac{n^\prime_{b}}{n_{b}Z_{b}}\right)=0
\end{multline}
\begin{multline} \label{eq:log_nc_prime}
(\log n_{c})^{\prime}=\left(\frac{C_{ca}}{Z_{a}}{n^\prime_{a}}+\frac{C_{cb}}{Z_{b}}{n^\prime_{b}}\right)\\
\times\frac{Z_{c}}{(n_{a}C_{ca}+n_{b}C_{cb})},
\end{multline}
where the collisionalities $C_{ca}$, $C_{cb}$ are defined as:
\begin{equation}
C_{ca}=Z_{c}^{2}Z_{a}^{2}\sqrt{\frac{m_{a}}{m_{a}+m_{c}}}\sqrt{\frac{1}{m_{c}}}\label{eq:C_ca}
\end{equation}
\begin{equation}
C_{cb}=Z_{c}^{2}Z_{b}^{2}\sqrt{\frac{m_{b}}{m_{b}+m_{c}}}\sqrt{\frac{1}{m_{c}}}\label{eq:C_cb}.
\end{equation}
Solving for $n_c$ in Eq.~(\ref{eq:log_nc_prime}) and integrating we arrive at the general result for the density of the trace species $c$:
\begin{equation}\label{eq:nc_ito_nanb}
n_{c}=n_{c_0}e^{\int_{x_0}^{x}\left(\frac{C_{ca}}{Z_{a}}{n^\prime_{a}}+\frac{C_{cb}}{Z_{b}}{n^\prime_{b}}\right)\frac{Z_{c}}{(n_{a}C_{ca}+n_{b}C_{cb})}dx}.
\end{equation}
We can solve Eq.~(\ref{eq:nc_ito_nanb}) analytically in the special case where we take the electron density to be flat. 
Invoke charge quasineutrality, expressed in Eq.~(\ref{eq:charge_imcomp}), which follows from $q_a \Gamma_{ab} + q_{b} \Gamma_{ba} = 0$, so that $Z_{a} n_{a} + Z_{b}n_{b}$ is not changed by that collisional interaction, to express Eq.~(\ref{eq:log_nc_prime}) in terms of $n_{a}$, and integrate:
\begin{equation}\label{eq:charge_imcomp}
n_{a}Z_{a}+n_{b}Z_{b}=K
\end{equation}
\begin{equation}
n_{a}^{\prime}Z_{a}+n_{b}^{\prime}Z_{b}=0.
\end{equation}
Here K is a constant. Then Eq.~(\ref{eq:log_nc_prime}) becomes:

\begin{multline} \label{eq:nc_prime_div_nc_1a}
\frac{n_{c}^{\prime}}{n_{c}}=\left[\frac{C_{ca}}{Z_{a}}n_{a}^{\prime}+\frac{C_{cb}}{Z_{b}}\left(-n_{a}^{\prime}\frac{Z_{a}}{Z_{b}}\right)\right]\\
\times\frac{Z_{c}}{n_{a}C_{ca}+\left(\frac{K-n_{a}Z_{a}}{Z_{b}}\right)C_{cb}}\end{multline}

\begin{multline} \label{eq:nc_prime_div_nc_1b}
\frac{n_{c}^{\prime}}{n_{c}}=n_{a}^{\prime}Z_{a}\left(\frac{C_{ca}}{Z_{a}^{2}}-\frac{C_{cb}}{Z_{b}^{2}}\right)\\
\times\frac{Z_{c}}{n_{a}C_{ca}+\left(\frac{K-n_{a}Z_{a}}{Z_{b}}\right)C_{cb}}.
\end{multline}
Then substitute $C_{ca}$ and $C_{cb}$, and express in terms of reduced mass, $\mu_{ac}={m_{a}m_{c}}/{(m_{a}+m_{c})}$, and simplify:
\begin{equation} \label{eq:nc_prime_div_nc_2}
\frac{n_{c}^{\prime}}{n_{c}}=\frac{Z_{a}Z_{c}\left(\sqrt{\mu_{ca}}-\sqrt{\mu_{cb}}\right)n_{a}^{\prime}}{n_{a}Z_{a}\left(Z_{a}\sqrt{\mu_{ca}}-Z_{b}\sqrt{\mu_{cb}}\right)+KZ_{b}\sqrt{\mu_{cb}}}.
\end{equation}
Then integrating, we obtain:
\begin{equation}\label{eq:n_c}
\frac{n_{c}}{n_{c_0}}=\left(\frac{n_{a}Z_{a}^{2}\sqrt{\mu_{ca}}+n_{b}Z_{b}^{2}\sqrt{\mu_{cb}}}{n_{a_0}Z_{a}^{2}\sqrt{\mu_{ca}}+n_{b_0}Z_{b}^{2}\sqrt{\mu_{cb}}}\right)^{\frac{Z_{c}\left(\sqrt{\mu_{ca}}-\sqrt{\mu_{cb}}\right)}{Z_{a}\sqrt{\mu_{ca}}-Z_{b}\sqrt{\mu_{cb}}}}
\end{equation}
Let:
\begin{equation}\label{eq:c0}
C_{0}\doteq\left(\frac{\sqrt{2}e^{4}\log\Lambda}{12\pi^{3/2}\epsilon_{0}^{2}}\right)\left(\frac{n_{0}}{m_{p}^{1/2}T_{0}^{3/2}\varOmega_{p0}}\right),
\end{equation}
where $\log\varLambda$ is the Coulomb logarithm. To find $n_{c}^{\prime}$ we solve for $n_{a}^{\prime}$ in Eq.~(\ref{eq:nanb}) as follows: 
\begin{equation}
\left(n_{a}n_{b}^{-\frac{Z_{a}}{Z_{b}}}\right)^{\prime}=-\frac{Z_{a}^{2}e^{2}{B^{2}}{\Gamma}_{ab}n_{b}^{-\frac{Z_{a}}{Z_{b}}-1}}{m_{a}{T}C_{0}C_{ab}}
\end{equation}
\begin{multline}
n_{a}^{\prime}=-\frac{{B}^{2}{\Gamma}_{ab}}{m_{a}{T}C_{0}Z_{a}Z_{b}e^{2}\sqrt{\frac{m_{b}}{m_{a}(m_{a}+m_{b})}}}\\
\times\frac{1}{n_{a}\frac{Z_{a}}{Z_{b}}+n_{b}\frac{Z_{b}}{Z_{a}}}.
\end{multline}
Then substitute this result for $n_{a}^{\prime}$ in Eq.~(\ref{eq:nc_prime_div_nc_2}), and use the following definitions:
\begin{align}
\mu_{r} &=\frac{\mu_{bc}}{\mu_{ac}}\\
\zeta^{-1} &=\frac{Z_{b}}{Z_{a}}.
\end{align}
Finally, we arrive at the expression:
\begin{multline} \label{eq:n_c_prime}
\frac{n_{c}^{\prime}}{n_{c}}=\left(\frac{{B}^{2}Z_{c}\Gamma_{ab}}{{T}C_{0}Z_{a}^{3}\sqrt{\mu_{ab}}}\right)\\
\times\frac{\left(\sqrt{\mu_{r}}-1\right)}{\Big(n_{a}+n_{b}\zeta^{-2}\sqrt{\mu_{r}}\Big)\Big(n_{a}+n_{b}\zeta^{-2}\Big)},
\end{multline}
where we have three free parameters: $\zeta$, $\mu_{r}$ and $n_{a}$ (since $n_{b}$ is constrained by Eq.~(\ref{eq:charge_imcomp}) given $n_{a}$). 
This expression is a key result. It describes how the trace species density gradient responds to other system parameters. In our setup, the sign of the right-hand-side of Eq.~(\ref{eq:n_c_prime})  depends only on the $\left(\sqrt{\mu_{r}}-1\right)$ term. Therefore, we conclude that $n_{c}^{\prime}>0$ if and only if $m_{b}>m_{a}$. In other words, the trace species tends to lump with the more massive bulk species irrespective of bulk species density or charge. 

We can simulate this scenario in MITNS for various ion species by adding a third ion species which is spatially uniform at $t=0$ and setting the trace density relative to the two bulk species. In Fig.~\ref{fig:FIG6_Contour_Carbon12} we use deuterium and tritium as the fuel species and Carbon-12 as the trace impurity, but similar simulations can be performed for the pB-11 fusion case. Our simulations of the density gradients agree with the analytic results in Eq~(\ref{eq:n_c_prime}), as shown in Fig.~\ref{fig:FIG7_Density_Carbon12}.

\begin{figure}
\includegraphics[width=\linewidth]{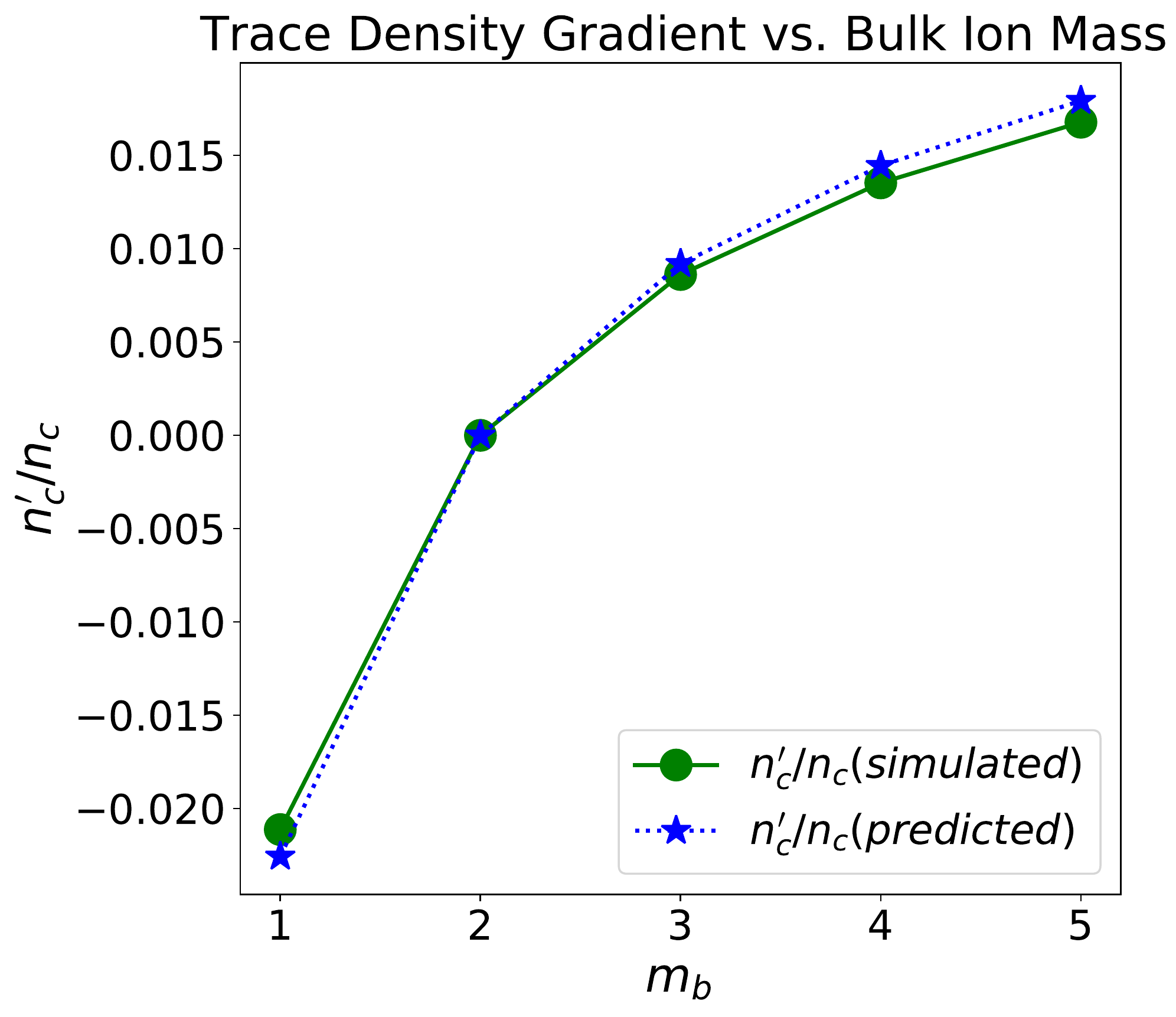}
\caption{\label{fig:FIG8_ChangingMass} As the mass of bulk ion species $b$ increases relative to the mass of bulk ion species $a$, the slope of the trace impurity density increases. In other words, the trace species density gradient is always in the direction of the higher mass. Here we set $m_a=2$ and $m_c=12$.}
\end{figure}

This surprising result is a consequence of the interplay of collisionality, the diamagnetic drift, and the density gradient of the bulk ion species in Eq.~(\ref{eq:nc_prime_div_nc_1b}). Highly charged ions collide a factor $Z^2$ more (Eqs.~(\ref{eq:C_ca}),~(\ref{eq:C_cb})), however, their diamagnetic drift per unit density gradient is slower by $1/Z$, Eq.~(\ref{eq:driftvel}), while the density gradient itself also scales inversely with $Z$ based on the quasineutrality of the plasma, see Eq.~(\ref{eq:charge_imcomp}). As predicted by Eq.~(\ref{eq:n_c_prime}), Fig.~\ref{fig:FIG8_ChangingMass} demonstrates that as the relative mass of species $b$ increases with respect to species $a$, the gradient of the trace density increases. The figure also shows that we find the same result using MITNS.

The extent and consequences of our findings depend on the species of ions present in the plasma device. For example, the impurities present in a fusion reaction would follow a different path within the reactor depending on the fuel. In the DT case, where the fuel is at the reactor's edges, the impurities would collect with the relatively more massive helium at the core. However, in the pB-11 case, the impurities will follow the boron ions to the edge of the reactor. 

A caveat to be considered is that, since the mass dependence in Eq.~(\ref{eq:n_c_prime}) is relatively weak, the Coulomb logarithm, which appears in the $C_0$ term, Eq.~(\ref{eq:c0}), may become important for low relative mass ratios of the background ion species. This is because Eq.~(\ref{eq:c0}) is an approximation that the Coulomb logarithm is the same for each species, though, in reality, a slightly different Coulomb logarithm enters each collision frequency. For mixed ion-ion collisions, the logarithm can be estimated as: 
\begin{multline}
   \lambda_{ic} = \lambda_{ci} = 23 - \ln\Bigg[\frac{Z_i Z_c(\mu_i+\mu_c)}{\mu_c T_c+\mu_c T_{i}}\\
   \times\left(\frac{n_{i}Z_i^{2}}{T_{i}}+\frac{n_{c}{Z_c}^2}{T_{c}}\right)^{1/2}\Bigg],
\end{multline}
where $\mu$ is the ratio of the ion mass to proton mass for ions $i$ and trace impurity $c$. For our case, the temperature is constant, and the trace species density $n_c$ is negligible. So the expression becomes:
\begin{equation}
    \lambda_{ic} = 23 - \ln\left[{Z_i}^{2} Z_c\left(\frac{n_i}{T_i^{3}}\right)^{1/2}\right],
\end{equation}
where $T$ is the temperature in $eV$ and $n_i$ is in $cm^{-3}$. However, this term can be safely ignored when $\left|\ln\left({Z_i}^{2}n_{i}^{1/2}T_{i}^{-3/2}\right)\right|\ll23$. Otherwise, it may have a significant impact on the outcome.

\section{\label{sec:Discussion_and_Conclusion}Discussion}

This work is the first generalization of classical impurity transport results, Eq.~(\ref{eqn:impurityPinch}) and (\ref{eqn:generalizedPinch}), to consider equilibria with steady-state particle fluxes. Unlike Eqs.~(\ref{eqn:impurityPinch}) and (\ref{eqn:generalizedPinch}), the finite-flux behavior depends on how many species are included, whereas Eqs.~(\ref{eqn:impurityPinch}) and (\ref{eqn:generalizedPinch}) must only be satisfied pairwise by all species in the plasma, no matter how many there are. Thus, we present several particular cases to understand and predict the behavior of impurities in a variety of situations. We demonstrate that the trace impurity density gradient in multi-ion-species plasmas tends to align with the more massive background species, which can be a significant consideration in real-world experiments and applications. For example, in a plasma-based mass separation centrifuge, trace species will separate from the lower-mass ions and will be pushed out with the heavier ions \cite{gueroult2014double,gueroult2015plasma}. Moreover, in controlled fusion, impurities can result from the plasma's interaction with the walls of the reactor, hence decreasing the temperature of the plasma and eroding the efficiency of the reactor.

Our results suggest that the high-mass-impurity alignment is beneficial in pB-11 fusion, as impurities are naturally expelled, but harmful in DT fusion, as they follow the helium to the core of the reactor. Furthermore, a high mass disparity between the background species will increase the trace ion density gradient but will not change its direction. A higher trace charge relative to the background species will also amplify this effect. Thus, the magnitude of the effect will be higher in the pB-11 fusion case than in DT fusion.

These results can be tested in magnetized plas-ma with a steady-state cross-field flow. A possible experiment setup would involve following the density evolution of a trace ion species in a mass separating device or a magnetically confined linear plasma device. Another possibility is feeding a trace amount of carbon-12 into a steady cross-field flow of D and T fuel. This might be done in a magnetized Z-pinch geometry.

\section*{Declaration of competing interest}
The authors declare that they have no known competing financial interests or personal relationships that could have appeared to influence the work reported in this paper.

\section*{Acknowledgments}
Elena Litvinova Mitra was supported by the Department of Energy for the Summer Undergraduate Laboratory Internship (SULI) program.  This work was additionally supported by DOE DE-AC02-09CH11466, NNSA 83228-10966 [Prime No. DOE (NNSA) DENA0003764], and NSF-PHY-1805316.

\bibliographystyle{apsrev4-1} 

\bibliography{MAIN_Trace_Dynamics.bib}

\end{document}